\documentstyle[epsfig,12pt]{article}

\def\reff#1{(\ref{#1})}

\def\beq{\begin{equation}}
\def\eeq#1{\label{#1}\end{equation}}

\def\dfrac#1#2{{\displaystyle\frac{#1}{#2}}}

\begin{document}
\begin{center}
{\bf Gauge choice and geodetic deflection in conformal gravity}
\end{center}
\vspace{1em}
\centerline {A. Edery$^1$, A. A. M\'ethot$^2$ and M. B. Paranjape$^3$}
\vspace{1em}
\begin{center}
{\it{$^1$}Center for High Energy Physics, Department of Physics,\\
McGill University,
3600 University Street,\\ Montr\'eal, Qu\'ebec, Canada, H3A 2T8}
\end{center}
\begin{center}
{\it{$^{2,3}$}Groupe de Physique des Particules, D\'epartement de
Physique,\\
Universit\'e de Montr\'eal,
C.P. 6128, succ. centre-ville,\\ Montr\'eal, Qu\'ebec, Canada, H3C
3J7}
\end{center}
\begin{center}
{\it{$^3$}National Center for Theoretical Sciences, Physics
Division,\\National Tsing Hua University\\ Hsinchu, Taiwan 300,
ROC}
\end{center}

\footnotetext[1]{edery@hep.physics.mcgill.ca}
\footnotetext[2]{methot@lps.umontreal.ca}
\footnotetext[3]{paranj@lps.umontreal.ca}

\vspace{3em}
\centerline{\bf Abstract}
\vspace{1em}

{\small Conformal gravity has been proposed as an alternative
theory of gravity which can account for flat galactic rotation
curves without recourse to copious quantities of dark matter.
However it was shown that for the usual choice of the metric, the
result is catastrophic for null or highly relativistic geodesics,
the effect is exactly the opposite yielding an effective repulsion
and less deflection in this case.  It is the point of this paper,
that any result for massive geodesics depends on the choice of
conformal gauge, in contradistinction to the case of null geodesics.
We show how it is possible to choose the gauge so that the theory
is attractive for all geodesics. } \vfill\eject

One of the most important problems facing astrophysics is that of
missing matter.   All observations indicate that particles,
massive or massless, are attracted more strongly by cosmological
sources than would be expected on the basis of the observable
(luminous) mass of the source, together with the use of Newton/Einstein
gravity\cite{review}.  There are two evident potential solutions to this
problem.  We can
postulate the existence of (copious) quantities of non-luminous
matter (dark matter), in just the right distribution, to account
for the excess attraction, or we can reject Einstein gravity and
propose an alternative theory of gravity\cite{km, Milgrom, Bekenstein,
Sanders} chosen to better reflect the
cosmological phenomenology.  Either solution implies a radical
departure from our present understanding of cosmology.  Conformal
gravity, in recent years, has been proposed as such an alternative theory
of gravitation.

Kazanas and Mannheim\cite{km} have had reasonable
success in fitting galactic rotation curves.  They consider the
spherically symmetric vacuum solution in conformal gravity
\beq
d\tau^{2}=B(r)\,dt^2 -
{dr^2\over B(r)} -r^{2}\left(d\theta^2 + \sin^{2}\theta\,d\varphi^{2}\right)
\eeq{ds2}
with
\beq
B(r)= 1-\dfrac{2\beta}{r} +\gamma\,r - kr^2
\eeq{solution}
where $\beta , \gamma$ and $k$ are constants.  $kr^2$ corresponds to a
cosmological
solution (conformal to a Robertson-Walker background) so that $k$ is chosen small enough not to
have any effect at galactic
scales while $\beta$ and $\gamma$ are chosen such that they do influence the
gravitation exactly at these scales.  Then for weak fields, at galactic
scales, the effective
Newtonian potential for non-relativistic matter is given by
\beq
\phi (r)= -\dfrac{2\beta}{r} +\gamma\,r .
\eeq{newton}
Clearly for $\gamma > 0$ this gives an additional attractive
linear potential.  Fixing $\gamma $ phenomenologically yields
reasonable fits to galactic rotation curves\cite{km}.  However, it
was recently pointed out that a positive sign for $\gamma$ yields
replusion for null and highly relativistic geodesics\cite{ep1} in blatant
contradiction with observation.  The deflection is given by
$\Delta\varphi ={4\beta\over r_0} - \gamma r_0$ where $r_0$ is the radius of
closest approach\cite{ep1}.  This seems to arrest the development of
the Kazanas-Mannheim program.

Our point in this paper is that it is possible to
circumvent this difficulty by an appropriate choice of the
conformal gauge.  The key observation is that while massless geodesics are
conformally invariant, massive geodesics are not.   A
massless geodesic satisfies
\beq
d\tau^{2}=0=B(r)\,dt^2 -
{dr^2\over B(r)} -r^{2}\left(d\theta^2 + \sin^{2}\theta\,d\varphi^{2}\right)
\eeq{null}
hence conformal rescaling $d\tau^{2}\rightarrow \Omega d\tau^{2}$,
evidently, has no effect on the geodesic equation.  Massive geodesics are,
however, sensitive to the conformal factor $\Omega$.

Take the sign of $\gamma$ in \reff{solution} to be negative,
$\gamma < 0$, so that
the conformal gravity implies additional attraction for
the null geodesics ($\Delta\varphi ={4\beta\over r_0} + |\gamma |r_0$).  The metric
\beq
d\tau^{2}=\Omega (r)\left( B(r)\,dt^2 -
{dr^2\over B(r)} -r^{2}\left(d\theta^2 +
\sin^{2}\theta\,d\varphi^{2}\right)\right)
\eeq{scaled}
will continue to be a  vacuum solution of conformal gravity with
the same massless geodesics.  We will show how to find $\Omega (r)$ so that
non-relativistic massive geodesics will also exhibit
additional attraction.  We will work to first order in the weak field
approximation, however, it is clear that our idea is not restricted
to this domain.  Massless geodesics are insensitive to the
conformal gauge whatever the gravitational field, while
massive geodesics are always sensitive to the conformal factor,
since the coupling of mass to the theory necessarily breaks the conformal
invariance.

To identify the effective Newtonian potential that
non-relativistic matter will feel, we must make a change of
coordinates bringing the metric back to the standard spherically
symmetric form.  We make the transformation
\beq
r^\prime
=r\sqrt{\Omega (r)}
\eeq{coordtrans}
which is imposed to yield
\beq
d\tau^{2}=B^\prime (r^\prime )\,dt^2 - {\,d{r^\prime }^2\over
B^\prime (r^\prime )} -{r^\prime}^2\left(d\theta^2 +
\sin^{2}\theta\,d\varphi^{2}\right)
\eeq{ds2prime}
with
\beq
B^\prime (r^\prime )= 1-\dfrac{2\beta}{r^\prime } -
\gamma\, r^\prime - k{r^\prime }^2,
\eeq{solnprime}
since $\gamma $ is now negative.
This yields the set of equations
\begin{eqnarray}
\Omega (r) B(r)&=&B^\prime (r^\prime )\label{eq1}\\
\dfrac{\Omega (r) dr^2}{B(r)}&=&\dfrac{d{r^\prime}^2}{ B^\prime (r^\prime )}\label{con1}.
\end{eqnarray}
These equations  can be easily solved
in first order, weak field perturbation theory. The weak field limit is phenomenologically, abundantly
justified.
We take $\Omega (r) =1+\psi (r)$ and assume that all $r$ dependent
functions are small compared to 1.  Then expanding \reff{eq1} to first order immediately yields
the solution
\beq
\psi (r) = -2\gamma r .
\eeq{11}
Using
\beq
{dr^\prime\over dr}={d\over dr}\sqrt{1-2\gamma r}r\approx {d\over dr}
(1-\gamma r)r = 1-2\gamma r
\eeq{12}
and using \reff{coordtrans} and noting that $r$ may be replaced with
$r^\prime$ in terms
that are already of first order, yields,
\begin{eqnarray}
{\Omega (r)dr^2\over B(r) }&\approx &{(1-2\gamma
r)dr^2\over 1 -{2\beta\over r} +\gamma
r -k r^2}\\
&\approx &
{d{r^\prime }^2\over 1 -{2\beta\over r^\prime} -\gamma r^\prime
-k{r^\prime}^2}
={d{r^\prime }^2\over B^\prime (r^\prime )}
\end{eqnarray}
verifying \reff{con1}.

Thus we see that the sign of $\gamma$ can be reversed with a
coordinate change and a concomitant change of the conformal gauge
in first order weak field perturbation.  The null geodesics are
insensitive to these transformations, indeed, if $r =r(\theta )$
is the original orbit equation for the null geodesic, then it is just
transformed to $r^\prime (\theta )= r^\prime (r(\theta ))$.  The
range of $\theta$ in the orbit equation, which determines the
scattering angle, is unchanged, hence neither is the scattering
angle or the deflection.  The sign of $\gamma$ being negative, this
choice corresponds to the null geodesics behaving as if there is additional
attraction in comparison to the usual Newton/Einstein case.  There are of course
local changes in the trajectory,
but these do not affect the global behaviour of the null orbits.

The massive geodesics on the other hand are sensitive to
conformal transformations.  Indeed the non-relativistic massive
geodesics for the metric \reff{ds2prime} are also deflected as if there
is additional attraction since the sign of $\gamma$ is negative, as
observed in \cite{km} and which consequently served as
the cornerstone of the conformal gravity program.

In conclusion, it is possible to choose a metric and conformal
gauge for which conformal gravity will yield additional attraction
above the Newton/Einstein result
for both massless and non-relativistic, massive particles.  Hence
the possibility that conformal gravity could give a solution of
the missing matter problem is not closed.  Much work should be
done, however, before embracing this theory as a viable
phenomenological alternative.  The exact scenario of the
spontaneous or explicit breaking of the conformal invariance must
be examined in detail.  Unless we understand the mechanism, the
theory will remain of only nominal theoretical interest.

\vspace{1em}
\centerline{Acknowledgements}
\vspace{1em}
We thank NSERC of Canada for partial financial support and the
Department of Physics, National Central University, Chungli, Taiwan, where
this paper was written.


\vspace{-1em}
\begin{thebibliography}{99}
\bibitem{review} For a review of the evidence see V. Trimble, {\sl ARA\& A}
{\bf 25}(1987) 425.
\bibitem{km}D. Kazanas and P.D. Mannheim, {\sl ApJ} {\bf 342} (1989) 635,
P.D. Mannheim, {\sl ApJ} {\bf 479} (1997) 659
and references therein, P.D. Mannheim, {\sl Gen. Rel. Grav.} {\bf 25} (1993)
697, P.D. Mannheim and D. Kazanas,
{\sl Gen. Rel. Grav.} {\bf 26} (1994) 337.
\bibitem{Milgrom}M. Milgrom, {\sl ApJ} {\bf 270} (1983) 365; {\sl ApJ} {\bf
270} (1983) 371; {\sl ApJ} {\bf 270} (1983) 384.
\bibitem{Bekenstein}J.D. Bekenstein and R.H. Sanders, {\sl ApJ} {\bf 429}
(1994) 480.
\bibitem{Sanders}R.H. Sanders, {\sl ApJ} {\bf 473} (1996) 117.
\bibitem{ep1}A. Edery and M. B. Paranjape,  Phys. Rev. D {\bf 58}
024011(1998).
\end{thebibliography}
\end{document}